\documentclass{ws-procs975x65}            
\def\figsubcap#1{\par\noindent\centering\footnotesize(#1)}
\begin{document}
\title{Searching for Radiative Neutrino Mass Generation\\
at the LHC}

\author{Raymond R. Volkas$^*$}

\address{
ARC Centre of Excellence for Particle Physics at the Terascale\\
School of Physics, The University of Melbourne\\
Victoria, 3010, Australia\\
$^*$E-mail: raymondv@unimelb.edu.au}

\begin{abstract}
In this talk, I describe the general characteristics of radiative neutrino mass models that can be probed at the LHC.  I then cover the specific constraints on a new, explicit model of this type.
\end{abstract}


\bodymatter

\section{See-saw versus radiative neutrino mass generation}\label{aba:sec1}

Many models of radiative neutrino mass are possible, some of which have been analysed in depth, while others have been examined only briefly, with quite a number yet to be explicitly written down let alone analysed.  A useful organising principle for Majorana neutrino mass models is to use standard model (SM) effective operators that violate lepton number conservation by two units ($\Delta L = 2$ operators) as a base from which to construct new theories.~\cite{Babu:2001ex,deGouvea:2007xp,Bonnet:2009ej,Angel:2012ug,Bonnet:2012kz,Angel:2013hla,Krauss:2013gy}

Restricting ourselves to operators containing SM fermions and a single Higgs doublet, they occur at odd mass dimensions, with the lowest order one occurring at $d=5$.  That operator is the famous Weinberg operator, which has the schematic structure $LLHH$, where $L$ is a lepton doublet and $H$ the Higgs doublet.  After electroweak symmetry breaking, this operator (actually a set of operators because of the family structure) directly induces a Majorana neutrino mass for left-handed (LH) neutrinos given by the see-saw formula,
\begin{equation}
m_\nu \sim \frac{ \langle H \rangle^2}{M},
\end{equation}
where $M$ is the scale of the new physics that gives rise to the operator in the low-energy limit.

Underlying renormalisable theories yielding $LLHH$ are constructed by ``opening up'' the operator.  The type-1,-2 and -3 see-saw models are the minimal, tree-level ways to open up $LLHH$.~\cite{Minkowski:1977sc, Yanagida:1980, Glashow:1979vf, Gell-Mann:1980vs, Mohapatra:1980ia, Magg:1980ut, Schechter:1980gr,Wetterich:1981bx, Lazarides:1980nt, Mohapatra:1980yp, Cheng:1980qt, Foot:1988aq}  By starting with the effective operator, one may systematically construct all of the minimal underlying models.

Other $\Delta L = 2$ effective operators have mass dimension seven or higher and feature other lepton and quark fields in their expression, except for those in the generalised Weinberg class $LLHH(\bar{H}H)^n$.~\cite{Bonnet:2009ej,Bonnet:2012kz,Krauss:2013gy}  These additional particles have to be closed off in loops to produce a Majorana neutrino mass self-energy graph from the operator.  Thus, theories based on $d \ge 7$ operators (apart from the generalised Weinberg class) necessarily produce radiative neutrino mass generation.  Underlying renormalisable theories can then be systematically constructed by opening up the operators in all possible ways, subject in practice to minimality assumptions, just as the three see-saw models may be derived from the Weinberg operator.

A list of all gauge invariant $\Delta L = 2$ operators at $d=5,7,9,11$, constructed from SM fermions and one Higgs doublet, was produced by Babu and Leung (BL).~\cite{Babu:2001ex}  The ninth operator in their list, $O_9 = LLLe^cLe^c$ (which has $d=9$), is the basis of the historically important Zee-Babu model of neutrino mass generation.~\cite{Zee:1985id, Babu:1988ki}  Two exotic scalars are introduced: an isosinglet, singly-charged state $h$ coupling to $LL$ and a doubly-charged, isosinglet $k$ coupling to $e^ce^c$.  Both carry two units of lepton number.  The cubic term $hhk$ combines with those interactions to induce $\Delta L = 2$ and produce Majorana neutrino mass at 2-loop order.  ATLAS has searched for $k$ through the same-sign dilepton channels $ee$, $e\mu$ and $\mu\mu$, deriving a lower bound of about 320 GeV on the mass~\cite{ATLAS:2012hi} (see also~\cite{Chatrchyan:2012ya}).

It is nice that the $\Delta L = 2$ operator perspective places the radiative neutrino mass models at one end of a systematic list of Majorana mass models, bookended by the tree-level see-saw models at $d=5$.  From the phenomenological viewpoint, radiative models are interesting because the mass scale of the new physics is lower than the favoured very high scale for see-saw models, and has more chance of encroaching into the LHC regime.

It is very much worth noting that upper bounds on the see-saw scales can be derived from naturalness considerations, because the new particles and interactions will destabilise the electroweak scale if the new physics occurs at too high a scale.  The type-1 upper bound was first computed by Vissani~\cite{Vissani:1997ys} to be $3 \times 10^7$ GeV, if the Higgs self-energy graph from neutrino Yukawa interactions is not to produce a correction to the $\mu^2$ parameter in the Higgs potential greater than $(1\ {\rm TeV})^2$.  This one-family result has recently been generalised to the realistic three-family case,~\cite{Clarke:2015gwa} producing the bounds
\begin{equation}
M_{N_1} \lesssim 4 \times 10^7\ {\rm GeV},\ \ M_{N_2} \lesssim 7 \times 10^7\ {\rm GeV},\ \ M_{N_3} \lesssim 3 \times 10^7\ {\rm GeV}\, \left(\frac{0.05\ {\rm eV}}{m_{\rm min}}\right),
\end{equation}
where $N_{1,2,3}$ are the three heavy neutral lepton mass eigenstates with $M_{N_1} \le M_{N_2} \le M_{N_3}$, and $m_{\rm min}$ is the minimum light neutrino mass.  It is interesting that these bounds imply that standard thermal, hierarchical leptogenesis~\cite{Fukugita:1986hr}, which requires sufficiently massive heavy neutral leptons, must involve some level of fine-tuning for the minimal type-1 see-saw model. This is true for all three cases: $N_1$, $N_2$ and $N_3$ leptogenesis. Naturalness, if successful leptogenesis is desired, may be restored by extending the theory.  An obvious example is the supersymmetric extension, and another is to add extra Higgs doublets, to divorce the leptogenesis parameter space from neutrino mass generation (one version of this is discussed in Ref.~\cite{Davoudiasl:2014pya}).

\section{Opening up $d=7$ operators}

Let us examine the $\Delta L = 2$, $d=7$ operators from the BL list as the next most complicated cases after the Weinberg operator.  The flavour contents are
\begin{equation}
O_2 = LLLe^cH,\ \ O_3 = LLQd^cH,\ \ O_4 = LL\bar{Q}\bar{u}^cH,\ \ O_8 = L\bar{e}^c\bar{u}^cd^cH,
\end{equation}
where the BL numbering scheme has been used, with $Q$ the quark doublet, $d$ the RH down quark and $u$ the RH up quark.  Flavour structures $O_{3,4}$ each yield two independent operators once weak-isospin index contraction possibilities are taken into account.  Operator $O_3$ is the basis of the pioneering 1-loop Zee model,~\cite{Zee:1980ai} while models constructed from $O_3$ and $O_8$ have been analysed in depth by Babu and Julio.~\cite{Babu:2010vp,Babu:2011vb}  In addition, $d=7$ contains the Weinberg-operator generalisation $O'_1 = LL\bar{H}HHH$.~\cite{Bonnet:2009ej,Krauss:2013gy}

\begin{figure}[h]%
\begin{center}
  \parbox{1.3in}{\includegraphics[width=1.2in]{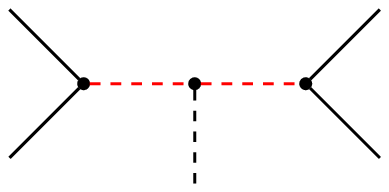}\figsubcap{a}}
  \hspace*{4pt}
  \parbox{1.3in}{\includegraphics[width=1.2in]{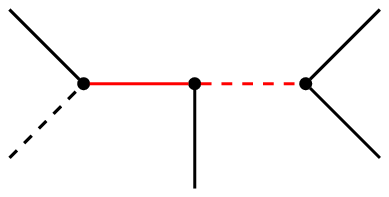}\figsubcap{b}}
  \hspace*{4pt}
  \parbox{1.3in}{\includegraphics[width=1.2in]{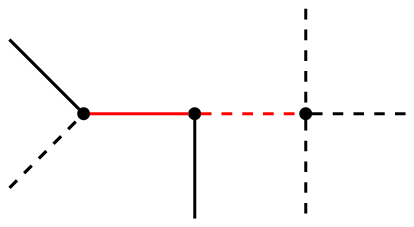}\figsubcap{c}}
  \caption{Diagram topologies for opening up $d=7$ operators. (a) Scalar-only. (b) Scalar plus fermion for $O_{2,3,4,8}$. (c) Scalar plus fermion for $O'_1$.}%
  \label{fig:topologies}
\end{center}
\end{figure}

We adopt the minimality assumption that the underlying renormalisable theories obtained from opening up these operators contain only exotic heavy scalars and vector-like fermions.  Table~\ref{tb:exoticscalarfermion} lists the quantum numbers under SU(3)$\times$SU(2)$\times$U(1) for scalar-only and scalar-fermion extensions.~\cite{Cai:2014kra}  The example in boldface will be analysed in detail below.  The diagram topologies are given in Fig.~\ref{fig:topologies}.

\begin{table}
\tbl{Quantum numbers of exotic scalars and fermions in underlying theories for the $d=7$ operators.}
{\begin{tabular}{@{}cccc@{}}
\toprule
Scalar & Scalar & Dirac fermion & Operator \\\colrule
(1,2,1/2) & (1,1,1) & & $O_{2,3,4}$ \\
(3,2,1/6) & (3,1,-1/3) & &  $O_{3,8}$ \\
(3,2,1/6) & (3,3,-1/3) & & $O_3$ \\
(1,1,1) & & (1,2,-3/2) & $O_{2}$ \\
(1,1,1) & & (3,2,-5/6) & $O_{3}$ \\
(1,1,1) & & (3,1,2/3) & $O_3$ \\
(3,2,1/6) & & (3,2,-5/6) & $O_3$ \\
{\bf (3,1,-1/3)} & & {\bf (3,2,-5/6)} & \boldmath$O_{3,8}$ \\
(3,3,-1/3) & & (3,2,-5/6) & $O_3$ \\
(3,2,1/6) & & (3,3,2/3) & $O_3$ \\
(1,1,1) & & (3,2,7/6) & $O_4$ \\
(1,1,1) & & (3,1,-1/3) & $O_4$ \\
(3,2,1/6) & & (3,2,7/6) & $O_8$ \\
(3,2,1/6) & & (1,2,-1/2) & $O_8$ \\
(1,4,3/2) & & (1,3,-1) & $O'_1$ \\\botrule
\end{tabular}}
\label{tb:exoticscalarfermion}
\end{table}

It is clear that searches for radiative neutrino mass models are included in general searches for exotic scalars and fermions.  Each specific model has a particular scalar-scalar or scalar-fermion pair.  In addition, the decay branching ratios for the exotics are typically quite constrained by the requirement to reproduce the correct neutrino masses and mixing angles.  This additional information must be used when interpreting the experimental constraints, and it constitutes a ``smoking gun'' for exotics that are responsible for radiative neutrino mass generation.

\section{Collider constraints on a new radiative model}

We now look at the boldfaced case in Table~\ref{tb:exoticscalarfermion}.~\cite{Cai:2014kra}  Our theory will produce $O_3$, with a subdominant $O_8$.  The exotic scalar $\phi$ and exotic fermion $\chi$ have the quantum numbers
\begin{equation}
\phi \sim (\bar{3},1,1/3),\qquad \chi \sim (3,2,-5/6).
\end{equation}
The Lagrangian is
\begin{eqnarray}
-{\cal L} & = & m_\phi^2 \phi^\dagger \phi + m_\chi \bar{\chi} \chi + \left( Y^{\bar{e}\bar{u}\phi}_{ij} \bar{e}_i \bar{u}_j \phi^\dagger + h.c. \right) \nonumber\\
& + & \left( Y^{LQ\phi}_{ij} L_i Q_j \phi + Y^{L\bar{\chi}\phi}_i L_i \bar{\chi} \phi^\dagger + Y^{\bar{d}\chi H}_i \bar{d}_i \chi H + h.c. \right)
\end{eqnarray}
where we use two-component notation for the fermions and $i,j$ are family indices.  The Yukawa coupling constants $Y^{\bar{e}\bar{u}\phi}$ play no role in neutrino mass generation and are thus set to zero for simplicity.  We also impose baryon-number conservation to forbid the $QQ\phi^\dagger$ and $\bar{d}\bar{u}\phi$ interactions permitted by the gauge symmetry.

\begin{figure}[h]
\begin{center}
\includegraphics[width=3in]{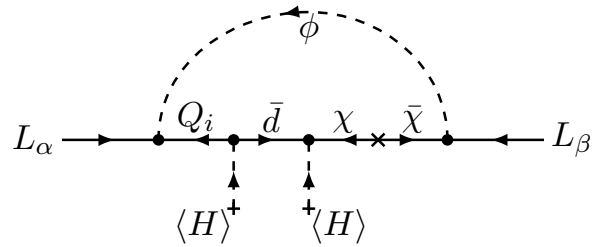}
\end{center}
\caption{Feynman diagram responsible for 1-loop generation of neutrino masses and mixings.}
\label{fig:NuMass}
\end{figure}

The diagram responsible for neutrino mass generation is given in Fig.~\ref{fig:NuMass}.  It is proportional to the down-type quark masses.  Barring unaesthetic hierarchies in the coupling constants, the terms proportional to the $b$-quark mass will dominate, which is what we will assume is the case.  Associated with that, for simplicity we switch off the mixing of $\chi$ with first- and second-generation quarks.

The neutrino mass matrix is then
\begin{equation}
(m_\nu)_{ij} = \frac{3}{16 \pi^2} \left( Y^{LQ\phi}_{i3} Y^{L\bar{\chi}\phi}_j + ( i \leftrightarrow j ) \right) m_{bB} \frac{m_b m_B}{m_\phi^2 - m_B^2} \ln \frac{m_B^2}{m_\phi^2}.
\label{eq:mnu-loop}
\end{equation}
The bottom quark mixes with the isospin $+1/2$ component of $\chi$ to form two mass eigenstates, $b$ and $B$, with masses $m_b$ and $m_B$ respectively.  The parameter $m_{bB} = Y^{\bar{d}\chi H}_3 v/\sqrt{2}$, where $\langle H^0 \rangle = v/\sqrt{2}$.  At this level of approximation, there are two massive neutrinos and a massless one.  In reality, the lightest neutrino will pick up a small mass, but its value is unimportant and will be neglected.  CP violating phases will also be set to zero for simplicity.

The rank-2 mass matrix may be expressed as the symmetrised outer product of two vectors $a_{\pm}$,
\begin{equation}
m_\nu = a_+ a_-^T + a_- a_+^T,
\label{eq:mnu-outer}
\end{equation}
where
\begin{equation}
a^{\rm NO}_{\pm} = \frac{\zeta^{\pm 1}}{\sqrt{2}} \left( \sqrt{m_2} u_2^* \pm i \sqrt{m_3} u_3^* \right),\quad
a^{\rm IO}_{\pm} = \frac{\zeta^{\pm 1}}{\sqrt{2}} \left( \sqrt{m_1} u_1^* \pm i \sqrt{m_2} u_2^* \right),
\end{equation}
with NO and IO denoting normal and inverted ordering, respectively.  The vectors $u_{1,2,3}$ are the columns of the PMNS matrix, and the complex number $\zeta$ is a Casas-Ibarra-like parameter, not determined by low-energy experiments.  Equating \ref{eq:mnu-loop} and \ref{eq:mnu-outer} constrains the parameter space to the region compatible with the neutrino oscillation results.

\begin{figure}[t]%
\begin{center}
  \parbox{2.1in}{\includegraphics[width=2in]{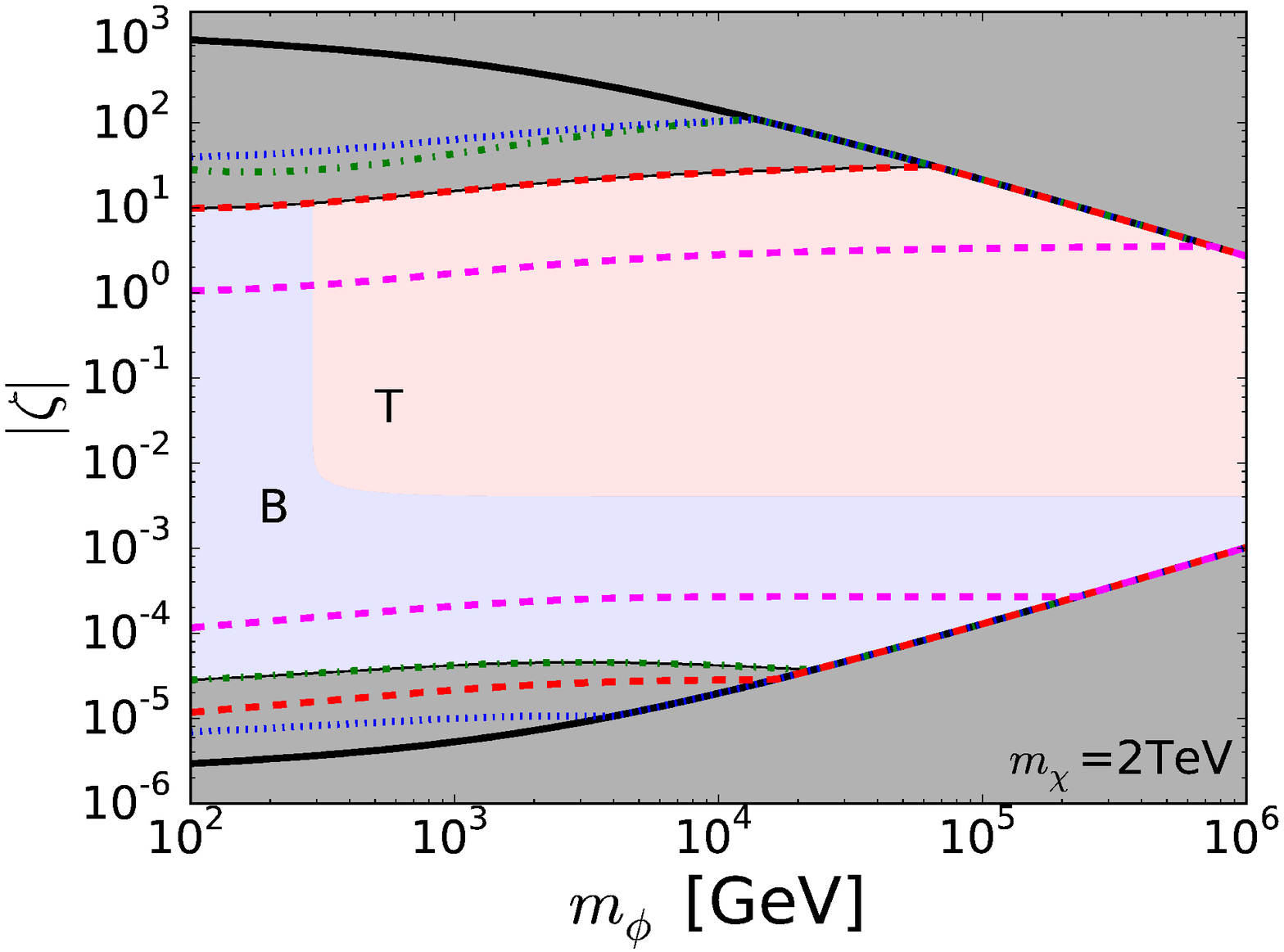}\figsubcap{a}}
  \hspace*{4pt}
  \parbox{2.1in}{\includegraphics[width=2in]{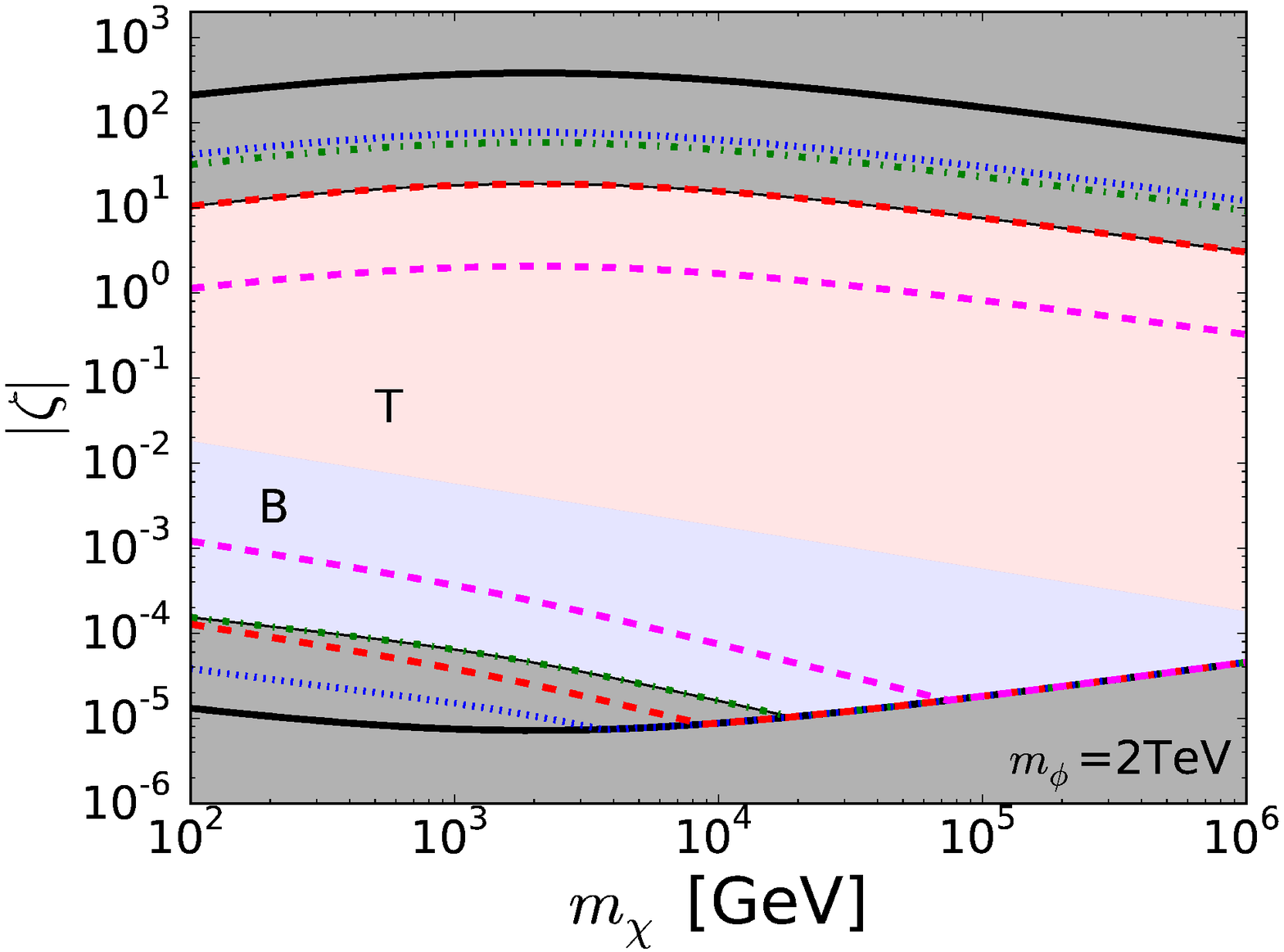}\figsubcap{b}}
  \caption{Lepton flavour violation bounds. (a) Allowed region in the $|\zeta|$, $m_\phi$ plane for $m_\chi = 2$ TeV. (b) Allowed region in the $|\zeta|$, $m_\chi$ plane for $m_\phi = 2$ TeV.}%
  \label{fig:LFV}
\end{center}
\end{figure}

The next set of constraints on the parameter space come from the lepton flavour violating processes $\mu \to e\gamma$, $\mu \to eee$ and $\mu N \to e N$.~\cite{Cai:2014kra} An example of the results is presented in Fig.~\ref{fig:LFV}. The green dash-dotted, blue dotted and red-dashed lines bound the regions corresponding to BR$(\mu \to e\gamma) < 5.7 \times 10^{-13}$, BR$(\mu \to eee) < 10^{-12}$ and BR$(\mu{\rm Au} \to e{\rm Au}) < 7 \times 10^{-13}$, respectively.  The grey region is thus excluded.  The magenta dashed line shows the reach of the $\mu{\rm Ti} \to e{\rm Ti}$ Mu2E and COMET experiments.  The allowed region is divided into $B$ and $T$ regions in which BR$(\phi \to b\nu)=1$ and BR$(\phi \to b\nu)<1$, respectively.

The vector-like fermion doublet $\chi$ consists of the charge $-1/3$ quark $B'$ which mixes with the bottom quark, and an exotic charge $-4/3$ quark $Y$.  CMS have searched for $B$-like particles, whose dominant decay modes are $B \to Z b$ and $B \to H b$ in our model with branching ratios that are strongly constrained by the need to fit the neutrino oscillation data.  Taking that into account, the CMS bound is $m_B \gtrsim 620$ GeV.~\cite{CMS:2013una,CMS:2012hfa,CMS:2013zea,Chatrchyan:2013uxa,Chatrchyan:2013wfa} No $Y$ search has been performed.

The other collider constraint comes from searches for leptoquark scalar $\phi$.  It is pair-produced at the LHC from gluon-gluon fusion and $q \bar{q}$ annihilation.  Since the production is through its colour charge, the rate depends only on $m_\phi$.  For example, for $m_\phi = 500 (600)$ GeV, the production cross-section $\sigma(pp \to \phi\phi)$ is 82(23.5) fb.

The main decay modes are into $Lt$ and $b\nu$, where $L = (e,\mu,\tau)$.  In the parameter region $m_{Y,B} \gg m_\phi$, so that $LY$ and $B\nu$ final states are kinematically forbidden, the partial decay rates are,
\begin{eqnarray}
\Gamma(\phi \to Lt) & = & \frac{m_\phi}{8\pi} \left| Y^{LQ\phi}_{L3} \right|^2 f(m_\phi,m_L,m_t),\nonumber\\
\Gamma(\phi \to \nu_L b) & \simeq & \frac{m_\phi}{8\pi} \left( \left| Y^{LQ\phi}_{L3} \cos\theta_2 \right|^2 + \left| Y^{L\bar{\chi}\phi}_{L} \sin\theta_1 \right|^2 \right) f(m_\phi,m_L,m_t),
\label{eq:decayrates}
\end{eqnarray}
where $f$ is given in Eqs.~4.16 and 4.21 of Ref.~\cite{Cai:2014kra}, and the mixing angles $\theta_{1,2}$ are defined through
\begin{equation}
\sin\theta_1 = \frac{m_{bB}m_\chi}{m^2_\chi - m^2_{b'}},\qquad \sin\theta_2 = \frac{m_{bB}m_{b'}}{m^2_\chi - m^2_{b'}},
\end{equation}
where $m_{b'} = y_b v/\sqrt{2}$ is the bottom quark mass in the absence of mixing with $B'$. The decay rates depend on the same Yukawa coupling constants that also contribute to neutrino mass generation, so are quite constrained, as well as $|\zeta|$.

\begin{figure}[h]
\begin{center}
\includegraphics[width=3in]{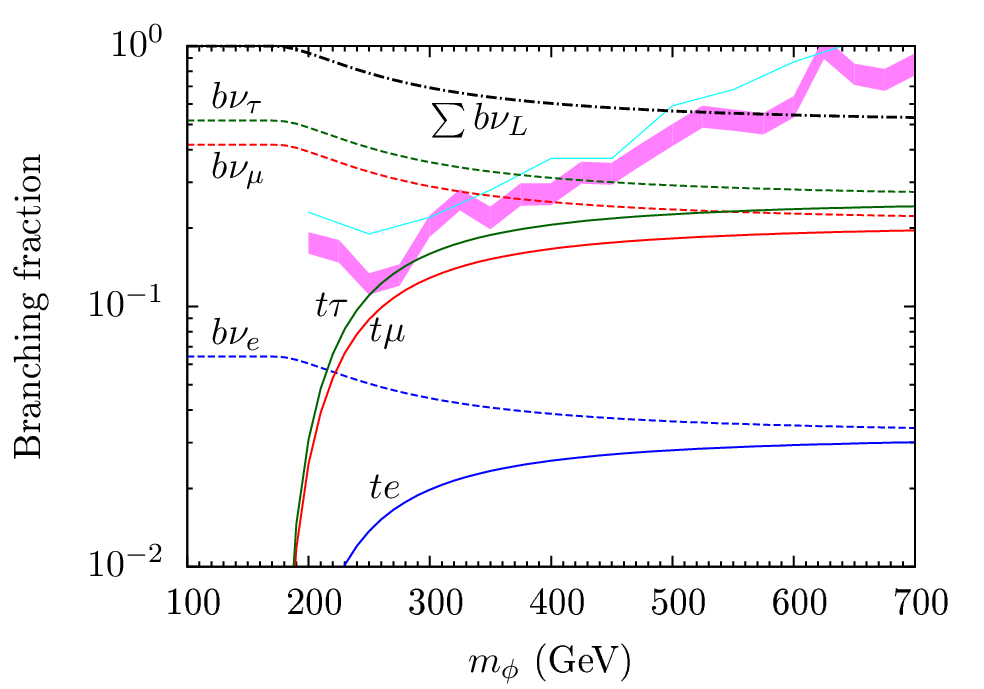}
\end{center}
\caption{Decay branching ratios of $\phi$ as a function of $m_\phi$ in region T. The curves with positive slope running to the top right show the ATLAS (thin line) and CMS (thick band) limits.}
\label{fig:regionT}
\end{figure}

Recall that the allowed regions $B$ and $T$ in Fig.~\ref{fig:LFV} correspond to regimes where BR$(\phi \to b\nu)=1$ and BR$(\phi \to b\nu)<1$, respectively.  In the $B$ region, the main signature is thus $pp \to \phi\phi \to bb + {\rm missing}\ E_T$, where sbottom pair searches apply.  The bound in this case is $m_\phi \gtrsim 730$ GeV at $95\%$ C.L.\cite{Aad2013ija,CMS-PAS-SUS-13-018}  The analysis for region $T$ requires recasting stop searches as well and is summarised in Fig.~\ref{fig:regionT}, which plots the various branching ratios as functions of $m_\phi$, as well as the limits from ATLAS and CMS.~\cite{Aad2013ija,CMS-PAS-SUS-13-018,ATLAS-CONF-2013-037,Chatrchyan2013xna,Aad2014qaa}  In numbers: The $m_\phi$ lower limit from CMS in the $bb + {\rm missing}\ E_T$ channel is in the range $520-600$ GeV. From the $(e,\mu) + {\rm missing}\ E_T + (b-){\rm jets}$, ATLAS sets a limit of approximately $580$ GeV.  Finally, the $(e,\mu)^+(e,\mu)^- + {\rm missing}\ E_T + {\rm jets}$ channel yields about a $620$ GeV limit from ATLAS data.

\section{Final remarks}

A systematic procedure for constructing radiative Majorana neutrino mass models has been developed from standard model $\Delta L = 2$ effective operators containing leptons, quarks and the Higgs doublet.  The LHC is a useful tool for searching for the exotic scalars and fermions that appear in these models, and specific bounds for a new model were summarised in this talk. The 13 TeV LHC will extend the reach of past searches. Some naturalness concerns for hierarchical, thermal leptogenesis in the minimal type-1 see-saw model were also mentioned in passing.

\section*{Acknowledgments}
This work was supported in part by the Australian Research Council.  I thank my coauthors P. W. Angel, Y. Cai, J. D. Clarke, R. Foot, N. L. Rodd and M. A. Schmidt.  I also thank K. S. Babu, A. de Gouv\^{e}a and W. Winter for discussions over the last few years on radiative neutrino mass generation.


\bibliographystyle{ws-procs975x65}
\bibliography{Volkas}

\end{document}